\documentstyle[prb,aps,amssymb,epsfig]{revtex}
\draft
\twocolumn
\title{Quantum Ising model in a transverse random field:\\ A
density-matrix renormalization group analysis}

\author{
A. Juozapavi\v{c}ius, S. Caprara,{\cite{sergio}} and A. Rosengren}
\address{
Department of Theoretical Physics, Royal Institute of Technology,
SE--100 44 Stockholm, Sweden}

\begin{document}
\maketitle
\begin{abstract}

The spin-1/2 quantum Ising chain in a transverse random magnetic field
is studied by means of the density-matrix renormalization group.
The system evolves from an ordered to a paramagnetic state as
the amplitude of the random field is increased. The dependence of the
magnetization on a uniform magnetic field in the $z$ direction and the
spontaneous magnetization as a function of the amplitude of the
transverse random magnetic field are determined. The behavior of the
spin-spin correlation function both above and at criticality is
studied.  The scaling laws for magnetization and correlation functions
are tested against previous numerical and renormalization-group
results.

\end{abstract}


\section{Introduction}

The spin-1/2 quantum Ising model in a transverse random magnetic field
(QIMRF) has been the subject of an increasing interest in recent
times. The main property of the QIMRF is that it undergoes a
disorder-driven quantum phase transition in the ground
state. Moreover, the critical point of the QIMRF is not an isolated
point of nonanalyticity, and the zero-field susceptibility diverges in
a wide region around the critical point. This behavior is due to the
existence of diluted domains of the wrong phase in the system. Such domains
are stabilized by the disorder.  This nonanalytical behavior is much
more pronounced in the quantum model than in its classical
counterpart, pointing out the difference between quantum and thermal
fluctuations in disordered systems.  More recently the QIMRF has been
obtained as an effective model for the one-dimensional Kondo lattice
model \cite{gulacsi} (KLM) by means of bosonization of the conduction
electrons. The critical behavior of the KLM has indeed features which
could be explained in terms of the corresponding QIMRF. \cite{capros}

Previous numerical results \cite{young} exploited the mapping of the
QIMRF in zero external uniform field onto a model of noninteracting
fermions. This mapping allows for a detailed study of correlation
functions, but is limited by the maximum size of the system that can
be handled. On the other hand, in the presence of an external uniform
field in the $z$ direction the model is not solvable. In this paper we
make use of the density-matrix renormalization-group (DMRG) approach
in its infinite-size version to overcome the limitations imposed by
the mapping to free fermions. By explicitly introducing an external
uniform field we are able to study the magnetization and extrapolate
the spontaneous magnetization, thus testing the scaling results and
determining the corrections beyond scaling.  Moreover our computing
time depends but linearly on the size of the system, for a given
accuracy.

In this paper we study the spin-1/2 QIMRF defined by the
Hamiltonian
\begin{equation}
\label{hamilton}
{\cal H}=-J\sum_i S_i^z S_{i+1}^z-\sum_i h_i S_i^x - h_z \sum_i S_i^z,
\end{equation}
where the $S_i^{\alpha}=\sigma_i^{\alpha}/2$, $\{\sigma_i^{\alpha} \}$
are Pauli spin matrices, $h_i$ are random-field values uniformly
distributed in the interval $-h_0<h_i<h_0$, and $h_z$ is a constant
field in the $z$ direction, added to produce spontaneous symmetry
breaking in our subsequent calculations. We take the units of energy
such that $J=1$. For the sake of definiteness we study the
ferromagnetic case $J>0$, to which the antiferromagnetic model can be
reduced by a gauge transformation on every second site ${\tilde
S}_i^{z}= (-1)^i S_i^{z}$, $i$ being the site index. Obviously in the
antiferromagnetic case $h_z$ has to be taken as a staggered field.
Moreover, a gauge transformation ${\tilde h}_i= {\textrm{sgn}}(h_i)
h_i$, ${\tilde S}_i^{x}= {\textrm{sgn}}(h_i) S_i^{x}$ can be performed
to make all $h_i$ positive.

\section{Previously known analytical results}

The QIMRF was studied analytically by McCoy and Wu,\cite{MW}
Shankar and Murthy,\cite{SM} and Fisher.\cite{fisher} Fisher, in
particular, makes use of the renormalization
group, to determine the critical properties of the model
defined by the Hamiltonian
\begin{equation}
{\cal H}^F=-\sum_i J_i^F \sigma_i^z \sigma_{i+1}^z - \sum_i h_i^F \sigma_i^x
- H \sum_i \sigma_i^z
\end{equation}
with all $J_i^F$ and $h_i^F$ positive random variables, drawn
independently from two distributions of densities $\pi (J), \rho (h)$.

The phase transition is found to occur at the point where
$\overline{\textrm{ln}h^F} = \overline{\textrm{ln}J^F}$, where the bar
denotes average over disorder. The distance from criticality is
defined as
\begin{equation}
\label{delta}
\delta \equiv \frac{\overline{\textrm{ln}h^F} -
\overline{\textrm{ln}J^F}}
{\textrm{var}(\textrm{ln}h^F)+\textrm{var}(\textrm{ln}J^F)}. 
\end{equation}

In Fisher's notation, the distributions chosen in this paper are
\begin{equation}
\label{distribution}
\pi (J)=\delta (J-1), \quad \rho (h)=\cases {h_0^{-1}&for $0<h<h_0$,\cr
0& otherwise.\cr}
\end{equation}
The relations between our and Fisher's variables are
$J=4J_i^F,~h_i=2h_i^F,~h_z=2H$.

Both in the ordered ($\delta<0$) and in the disordered ($\delta>0$)
phase a Griffiths\cite{griffiths} region does exist in which the
magnetization is nonanalytical at $H=0$. The weakly ordered Griffiths
``phase'' extends over the region where $\textrm{min} \{ J_i^F \} <
\textrm{max} \{ h_j^F\}$, $\delta <0$.  The weakly disordered
Griffiths ``phase'' extends over the region where $\textrm{max} \{
J_i^F \} > \textrm{min} \{ h_j^F \}$, $\delta >0$.

Fisher has found the exact critical scaling function of the
magnetization and he showed in particular that the magnetization at
the critical point behaves as
\begin{equation}
\label{five}
M\left(\delta =0,H\right) \sim \left[ \textrm{ln} \left( D_h/H\right) 
\right]^{\phi-2},
\end{equation}
where $\phi= (1+\sqrt{5})/2=1.618\,033\ldots$~ is the golden mean
and $D_h$ is some nonuniversal
scale factor. The spontaneous magnetization in the ordered phase depends
on the distance from criticality according to
\begin{equation}
\label{six}
M\left( \delta<0,H=0 \right) \sim (-\delta)^{2-\phi},
\end{equation}
for $\delta \rightarrow 0^-$. Further away from the critical point, in
the weakly ordered phase ($\delta<0$), for small $H$
\begin{equation}
\label{seven}
M\left( H \right)-M_0 \sim \left(H\right)^{1/(1+z)}
\left[ \textrm{ln}\left( D/H\right) \right]^x,
\end{equation}
where $M_0$ is the spontaneous magnetization, the exponent $x$ cannot
be determined within renormalization group, $D$ is a nonuniversal
scale factor, and
\begin{equation}
\label{nine}
z \approx \frac{1}{2|\delta|}.
\end{equation}

In the weakly disordered phase ($\delta>0$), instead
\begin{equation}
\label{eight}
M\left( H \right) \sim \delta^{3-\phi} \left(H \right)^{1/z} 
\left[ \textrm{ln}\left( D'/H\right)
 \right]^{1+1/z},
\end{equation}
where $D'$ is another nonuniversal scale factor, and $H\ll D'$.

The two-point spin-spin correlation function 
\begin{equation}
C(r)= \langle \sigma_0^z \sigma_r^z \rangle
\end{equation}
has the typical behavior
\begin{equation}
\label{lnc}
-\textrm{ln}C(r) \sim \sqrt{r}
\end{equation}
at the critical point $\delta=0$ in the limit of large $r$. The
$\langle \cdots \rangle$ denotes the quantum average in the ground
state. In the disordered phase $\delta>0$, $C(r)$ should obey the
scaling equation
\begin{equation}
\label{scaled}
\overline{\textrm{ln} \frac{C(r;\delta)}{C(r;\delta=0)}} =
\textrm{ln} \tilde{C}(r/\tilde{\xi}) \approx -r/\tilde{\xi},
\end{equation}
for $r \gg \tilde{\xi}$, where the typical correlation length
$\tilde{\xi}$ has the behavior
\begin{equation}
\tilde{\xi} \sim \frac{1}{\delta^{\tilde{\nu}}},
\end{equation}
with $\tilde{\nu}=1$.  Young and Rieger\cite{young} have, by mapping
the model in Eq.~(\ref{hamilton}) for $h_z=0$ onto a model of free
fermions, numerically calculated the typical correlation functions on
finite-size chains. They obtain that their data scale best for
$\tilde{\nu}=1.1$.

\section{DMRG algorithm}

The DMRG algorithm introduced by White\cite{white} is used with minor
changes. The random field breaks the symmetry of the chain under
reflection and separate density matrices are needed for the left and
the right subsystems. At each step of renormalization a new block is
defined by truncating the Hilbert space of the left and right
subsystems onto the eigenstates of the density matrix of the
subsystem which correspond to the largest eigenvalues. In our
calculation the minimum number of states kept is 8 at each iteration
step. However, if a degeneracy occurs in the density matrix, then all
the degenerate eigenstates are kept, a procedure which significantly
decreases the truncation error \cite{white}.  Even at the point of the
phase transition we found a relative truncation error of the order
$10^{-10}$, which makes our results very accurate.

The DMRG algorithm is used to calculate the average magnetization of
the chain and the spin-spin correlation functions at some given value
of the amplitude of the random field. In the first case the average of
the $z$ component of the spin operator of the full system is
calculated after 50 DMRG steps (the system is then composed of 102
sites). The value is averaged over a large number (usually 1000,
sometimes up to 6000, depending on the values of $h_z$) of different
random-field configurations. In the second case $N=20-40$ DMRG steps
are ignored to make the system large enough so that the edges would
have a small influence on the innermost spins. One spin in the middle
of the lattice is marked $a$ and the matrix of its $z$ component is
$S_a^z$ at the step $N+1$. The chain can be represented as $B_LasB_R$,
where $B_L$ denotes the left block of spins, $B_R$ the right block and
$s$ is another spin site in the middle of the lattice. The correlation
function of the $a$ spin with itself is trivially calculated as
$C(0)=\langle \delta_L \otimes (S_a^z)^2 \otimes \delta_s \otimes
\delta_R \rangle=1/4$. The $\delta$'s are unit matrices of the
corresponding parts of the system. At the step $N+2$ the left
subsystem $B_La$ is renormalized defining a new block $B'_L$,
according to White's procedure;\cite{white} in our procedure we
cannot, however, exploit the symmetry of the system under reflection,
so that the right subsystem has to be renormalized separately,
yielding a new block $B'_R$. Finally two more sites are inserted
between the new blocks. The system looks now like $B'_LbsB'_R$, where
the left one of the two new sites is called $b$. The spin $a$ is now
the rightmost spin in the block $B'_L$, so that its spin matrix is no
longer equal to the simple $2\times 2$ spin matrix, but it is some
matrix $S_L^z(1)$ of dimension $m\times m$, obtained by truncating the
product $\delta_L \otimes S_a^z$. Here $m$ is the number of states
retained in the block at each step (8 or more). The number $1$ in the
parentheses denotes the distance of the spin $a$ from the right edge
of the $B_L$. The correlation function between the spins $a$ and $b$
is $C(1)=\langle S_L^z(1) \otimes S_b^z \otimes \delta_s \otimes
\delta_B' \rangle$. After each RG step the spin $a$ ``moves'' deeper
into the block $B_L$ and a correlation function between spins that are
farther apart is calculated according to $C(r)=\langle S_L^z(r)
\otimes S_b^z \otimes \delta_s \otimes \delta_B' \rangle$.  We repeat
the procedure until a sufficient number (50) of values of $C(r)$ is
obtained.  All of them are averaged over a large number of random
field configurations (3000--8000).

It should be mentioned that the magnetization must be calculated at
nonzero (though very small) $h_z$ to have spontaneous symmetry
breaking.  This small additional field makes the magnetization larger
than the spontaneous magnetization $M_0=M(h_z=0)$. Several runs at
different $h_z$ are needed to extrapolate $M_0$. We shall return
to this point later.  On the other hand, as usual, the correlation
functions are directly calculated at $h_z=0$.

\section{Numerical results and discussion}

We carefully tested the reliability and accuracy of our algorithm in
the two exactly solvable limits of Hamiltonian (\ref{hamilton}), i.e.,
$h_0=0$ (pure Ising model) and $J=0$ (noninteracting spins in a random
field). In the first case we reproduced the exact ground-state
wavefunction $|\Phi\rangle = \prod_i |\uparrow_i\rangle$ (for $h_z\to
0^+$), the ground-state energy $E_0=-J(N-1)$, where $N$ is the number
of sites, and the correlation function $C(r)=1/4,~ \forall r$. In the
second case we checked that for a given random-field configuration,
the ground-state wavefunction is $|\Phi\rangle=2^{-N/2}\prod_i
\left[|\uparrow_i\rangle + {\textrm{sgn}}(h_i)
|\downarrow_i\rangle\right]$ with a ground-state energy
$E_0=-1/2\sum_i |h_i|$. We also reproduced the correct average of the
ground-state energy and correlation functions over disorder.

\begin{figure}[h]
\epsfxsize=7cm
\centerline{\epsfbox{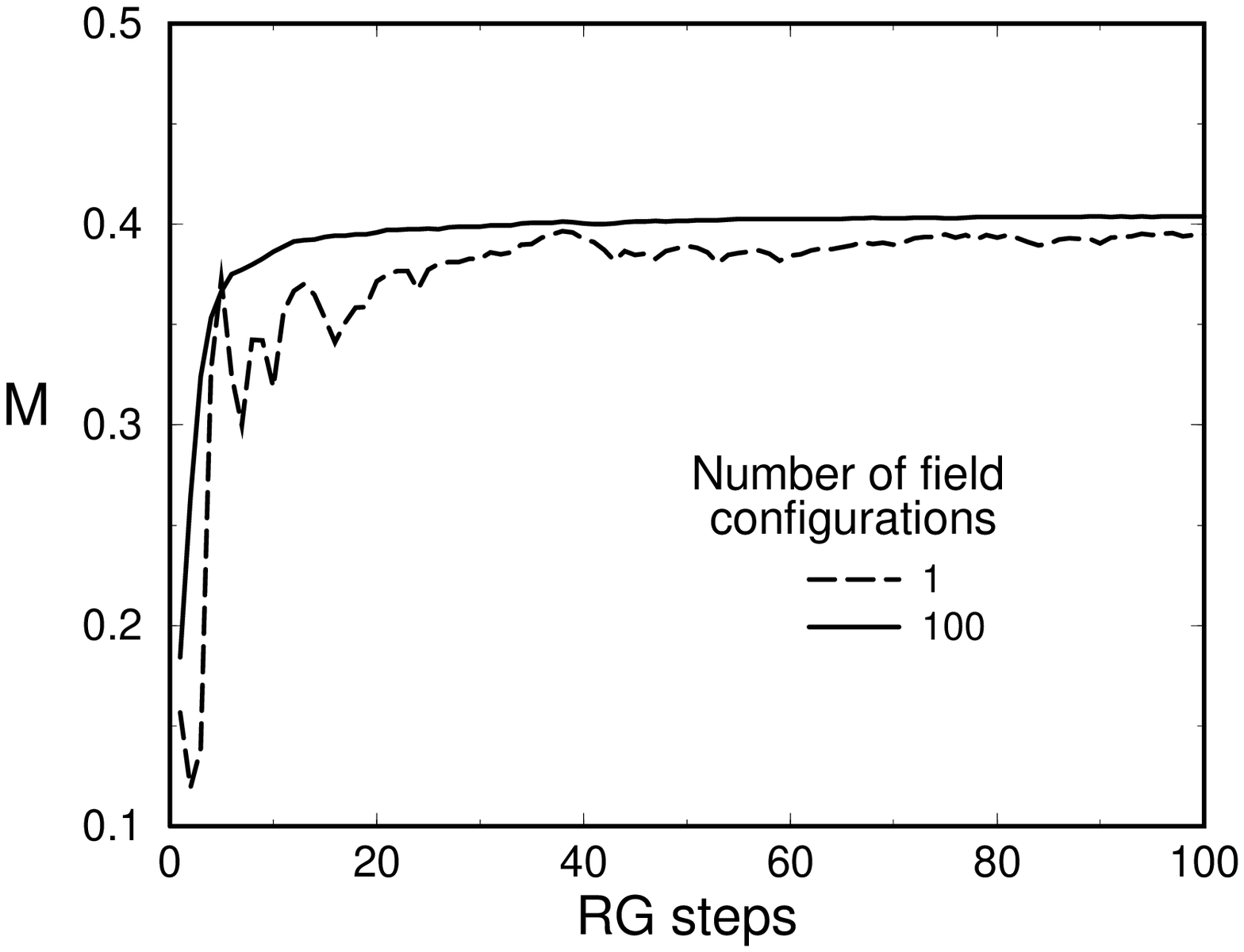}}
\caption{ The magnetization dependence on the system size. All our
further calculations are interrupted after 50 RG steps, when the
magnetization has saturated.}
\label{fig1}
\end{figure}

\begin{figure}[h]
\epsfxsize=7cm
\centerline{\epsfbox{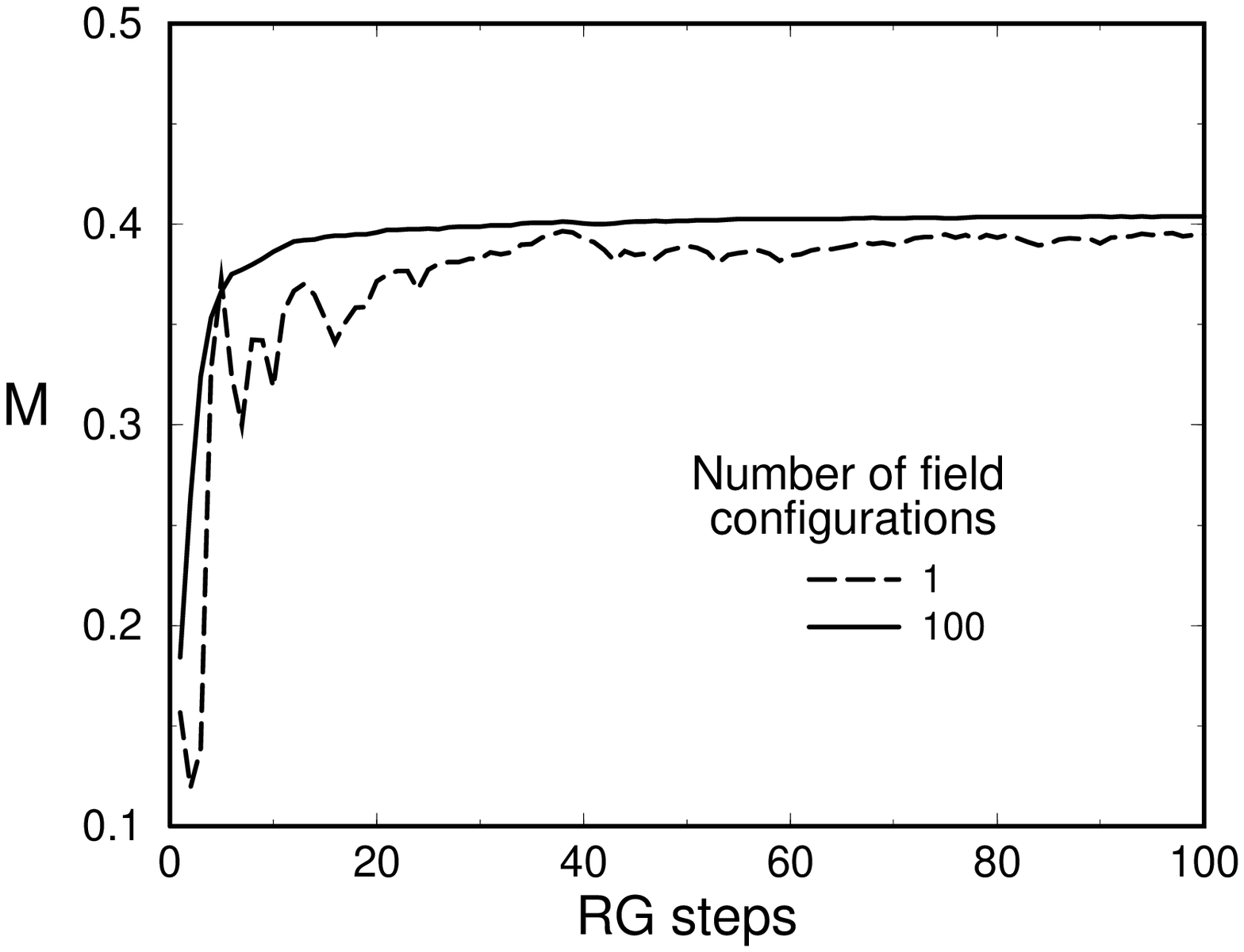}}
\caption{ 
The dependence of the magnetization $M$ and the estimated error of the
mean $\sigma_\mu$ on the number of random magnetic field
configurations $N_c$. Here $\sigma_\mu=\sqrt{\sum_i
(M_i-\overline{M})^2}/\sqrt{N_c(N_c-1)}$, where a bar denotes the
average over configurations.
}
\label{fig2}
\end{figure}

The dependence of the calculated quantities both on system size
(i.e., on the number of RG steps) and on the number of random-field
configurations has to be carefully discussed. A sufficiently large
number of RG steps should be performed to eliminate the effects of the
edges of the lattice. As is seen in Fig.~\ref{fig1}, the magnetization
is an increasing function of the size of the chain.  The effect in
this case becomes typically negligible after 50 RG
steps. Figure \ref{fig2} shows that oscillations of the magnetization as
well as the estimated error of the mean decrease quite rapidly with
increasing number of field configurations. Usually $1000$
configurations are enough for the relative error to become $1-2$~\%
though a larger number of them is required at small external fields.
Again we point out that in all our calculations the relative
truncation error was of the order 10$^{-10}$, so that the errors that
affect our data are purely statistical, the convergence to the
infinite-size limit being controlled by the number of RG steps, and
the accuracy of our averages being determined by the number of
random-field configurations.

\begin{figure}[h]
\epsfxsize=7cm
\centerline{\epsfbox{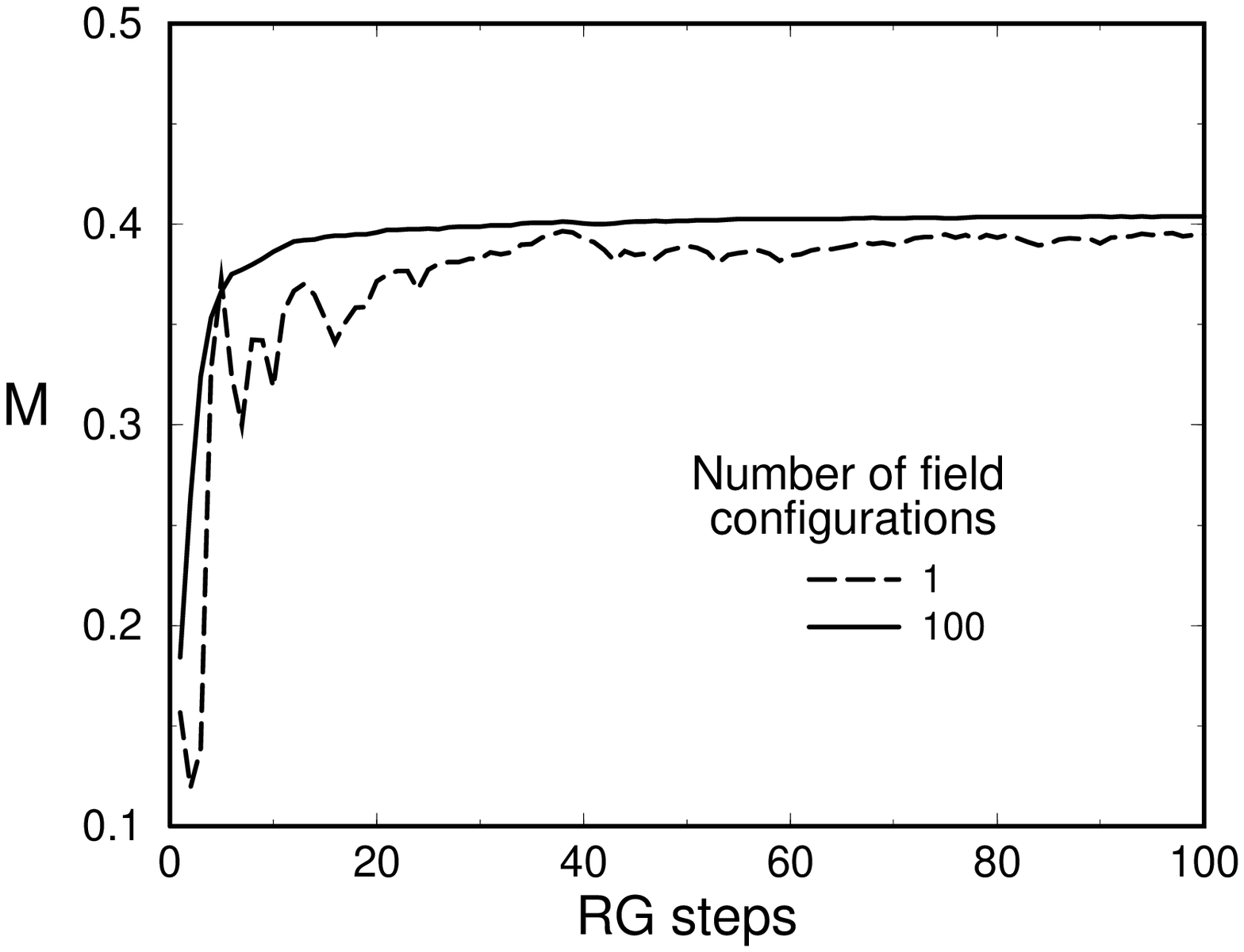}}
\caption{
The magnetization dependence on the amplitude of the random magnetic
field at the small ordering field $H=h_z/2=5\times 10^{-6}$ and
$H=0$. The error bars are smaller than the symbols. The data are
averaged over 1000 samples. The solid line is a fit according to
Eq.~(\ref{six}).}
\label{fig3}
\end{figure}

The magnetization dependence on the amplitude of the random magnetic
field $h_0$ at $h_z=1\times 10^{-5}$ is shown in Fig.~\ref{fig3}. The
system is in a ferromagnetic state at small fields and $M=1/2$. The
magnetization decreases with increasing $h_0$ and the system is driven
towards a paramagnetic state. Equation (\ref{delta}) predicts the
phase transition to occur at $h_0 =e/2 =1.3591 \ldots$~ since the
distribution (\ref{distribution}) gives in our case
\begin{equation}
\delta=\textrm{ln} \frac{2h_0}{e},
\end{equation}
once the mapping to Fisher's variables is taken into account. The
critical point is fairly well reproduced by our extrapolation.

The spontaneous magnetization $M(h_z=0)$ decreases according to
Eq.~(\ref{six}). The solid line is the best fit of the form
$M_0=a(-\delta)^{\beta}$ with $a=0.56 \pm 0.02$ and $\beta = 0.40 \pm
0.02$. The exponent $\beta$ agrees well with the predicted value
$2-\phi= (3-\sqrt{5})/2 \approx 0.382$. The values of $M_0$
where obtained by fitting Eq.~(\ref{seven}). In our fit we find that
the exponent $x$ is negligibly small.

\begin{figure}[h]
\epsfxsize=7cm
\centerline{\epsfbox{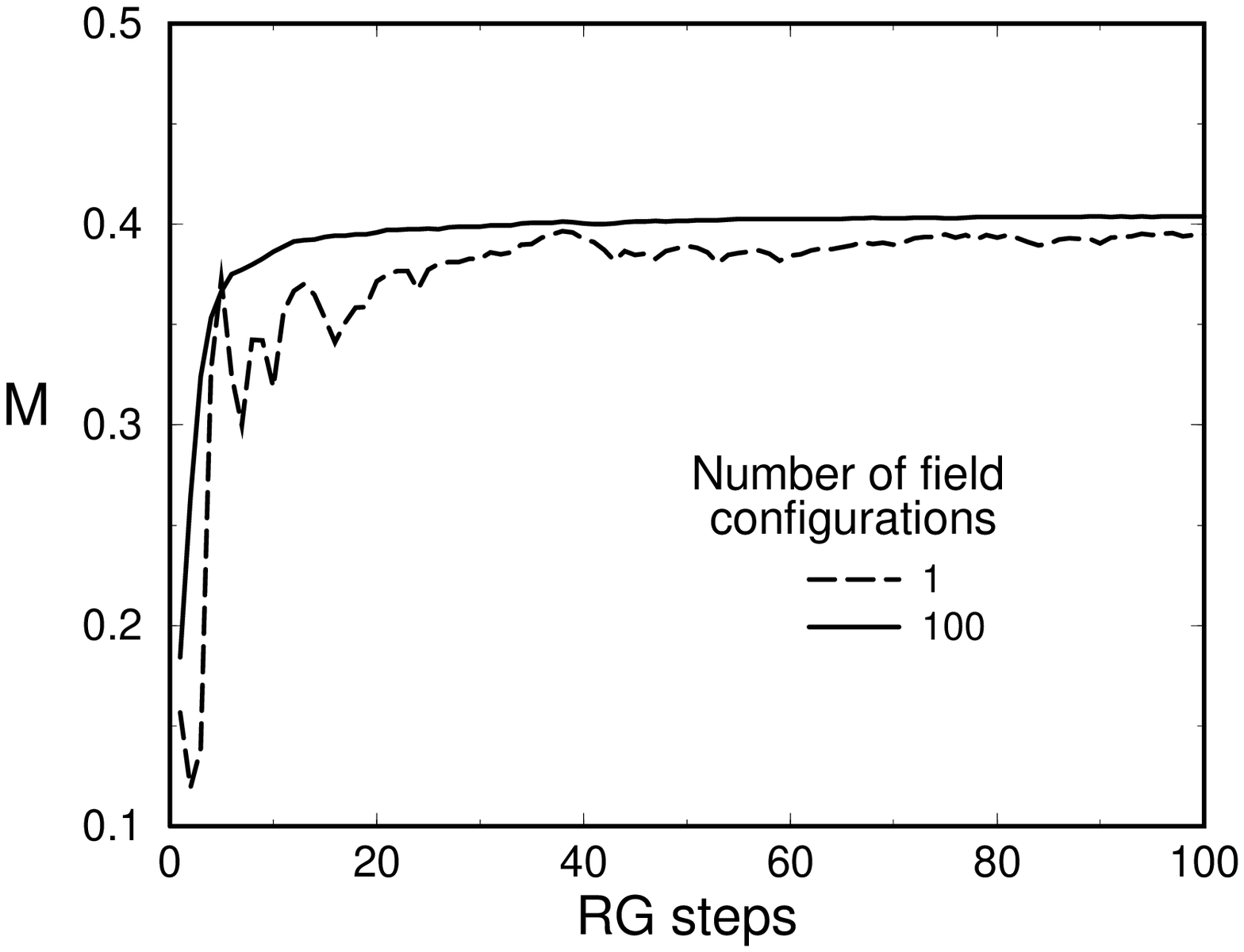}}
\caption{
The magnetization dependence on the ordering field at criticality. The
solid line is a fit of the form (\ref{five}).}
\label{fig4}
\end{figure}

The magnetization as a function of the external uniform field at the
critical point is plotted in Fig.~\ref{fig4}. The result agrees with
Eq.~(\ref{five}).  The solid line is a fit of the form $M(H)=c\left[
\ln {(D_h/H)}\right]^{\phi-2}$ with $c=0.34\pm 0.01$ and
$D_h=0.024\pm0.005$.

\begin{figure}[h]
\epsfxsize=7cm
\centerline{\epsfbox{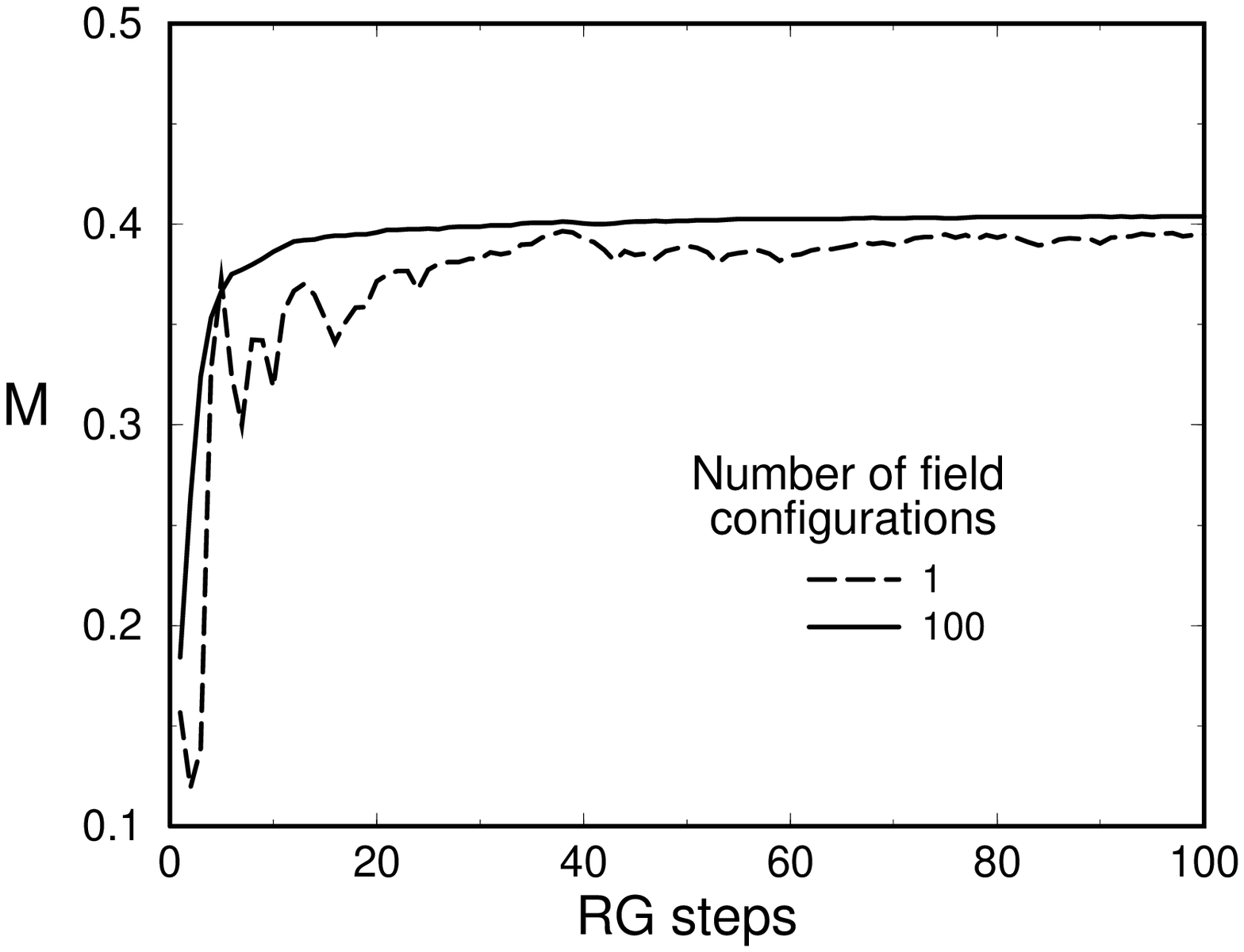}}
\caption{
The dependence of the inverse of the exponent $z$ on the distance from
criticality. The solid line is the theoretical behavior of $1/z$ for
small $\delta$ values, according to Eq.~(\ref{nine}).}
\label{fig5}
\end{figure}

We also addressed the issue of determining the deviations of the
exponent $z$ from its scaling value $1/(2|\delta|)$. By fitting
Eq.~(\ref{seven}) and Eq.~(\ref{eight}) for $\delta<0$ and $\delta>0$,
respectively, we obtained the $1/z$ as a function of $2|\delta|$,
plotted in Fig.~\ref{fig5}.  The exponent $z$ is smaller than its
scaling value in the ordered phase, and larger in the disordered
phase.

\begin{figure}[h]
\epsfxsize=7cm
\centerline{\epsfbox{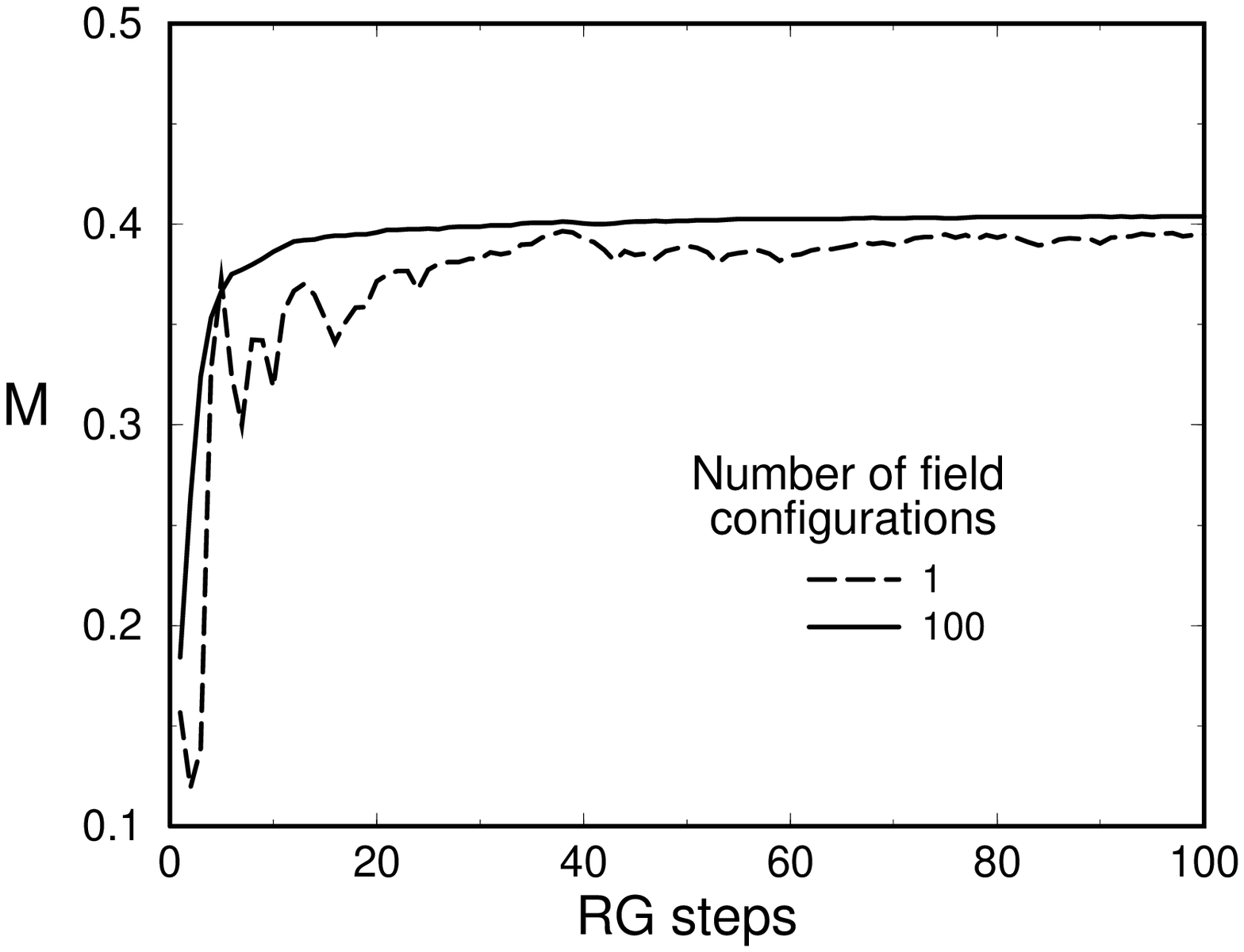}}
\caption{
The average of the logarithm of the correlation function at the
critical point. The straight solid line is a fit to the data with $20
\le r \le 50$, the slope of $0.96 \pm 0.01$, and the intercept of
$0.52 \pm 0.01$. The logarithm of the correlation function was
averaged over 8000 random field configurations.}
\label{fig6}
\end{figure}

The average of the logarithm of the two-point spin-spin correlation
function at the critical point is shown in Fig.~\ref{fig6}. According
to Eq.~(\ref{lnc}), $\overline{\textrm{ln}C}$ should be proportional
\noindent
to the square root of $r$ for large $r$. The solid straight line is a
fit to the data with $20 \le r \le 50$ and has a slope $-0.96 \pm
0.01$ and an intercept $0.52\pm 0.01$. Our result is thus
\begin{equation}
-\overline{ \textrm{ln}C} = 0.96(1)\ r^{1/2}-0.52(1),
\end{equation}
where the figures in the parentheses denote the standard error. 

\begin{figure}[h]
\epsfxsize=7cm
\centerline{\epsfbox{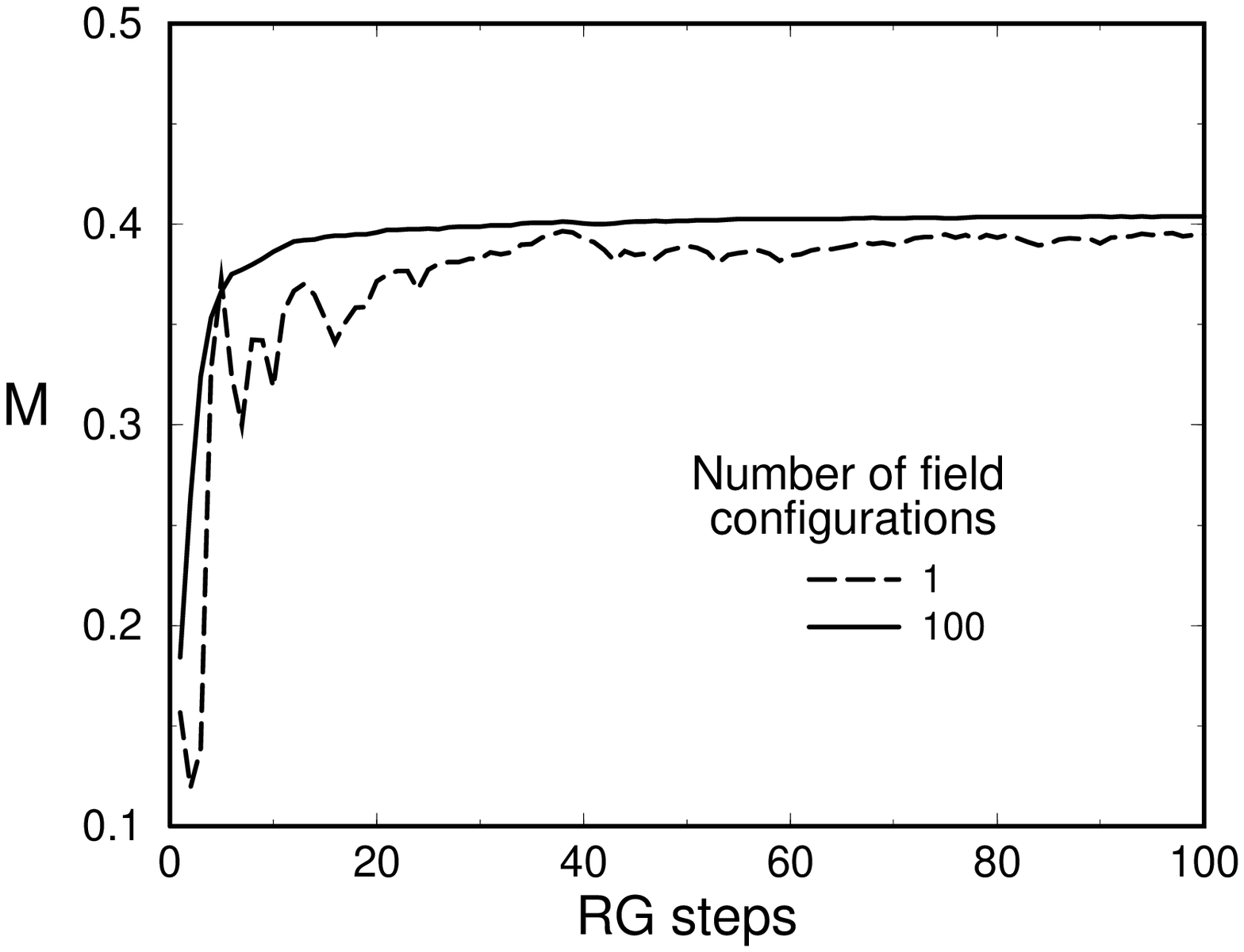}}
\caption{
A scaling plot of the average of the logarithm of the correlation
function in the disordered phase with the exponent $\tilde{\nu}=1.04$.}
\label{fig7}
\end{figure}

Our best scaling plot of the logarithm of the correlation function in
the disordered phase is shown in Fig.~\ref{fig7}.  The data for
different $\delta$ collapse on a straight line, of slope $\approx -1$, when
the exponent is $\tilde{\nu}=1.04$ which is slightly larger than the
scaling value, predicted by Fisher,\cite{fisher} $\tilde{\nu}=1$, but
smaller than the exponent determined by Young and Rieger,\cite{young}
which obtained the best scaling at $\tilde{\nu}=1.1$.
That the slope of the line is $\approx -1$ lends strong support to the
suggestion by Fisher, mentioned in Ref.~\onlinecite{young}, that
$\tilde{\xi}^{-1}=\overline{\textrm{ln}h^F} -
\overline{\textrm{ln}J^F}$, which for the distribution (4) gives that
$\tilde{\xi}^{-1}=\delta$ or $\ln \tilde{C}(r/\tilde{\xi})\approx
-\delta r$ for $r \gg\tilde{\xi}$. 

\section{Conclusions}

In summary, we have analyzed some properties of the spin-1/2 quantum
Ising model in a transverse random magnetic field (QIMRF) by means of
the density-matrix renormalization group (DMRG). This method has two
advantages with respect to other numerical methods. First of all, the
computing time, for a given accuracy, scales linearly with the size of
the system. Thus there is no severe limitation in increasing the size
of the system. Our results show that typically the algorithm converges
to the infinite-size limit after 50 -- 100 DMRG steps, i.e., when the
system is composed of 100 -- 200 sites.  Second, an external uniform
field in the $z$ direction could be explicitly added, allowing us {\it
to determine the scaling laws for the magnetization}, both as a
function of the external field and of the distance from
criticality. This is the most relevant aspect of our approach. Indeed
the introduction of the external field spoils the integrability which
was exploited in Ref.~\onlinecite{young}, but is absolutely harmless
from the point of view of the DMRG, the only effect being that a large
number of random-field configurations is needed for a given accuracy.

It is evident that once the solvability of the model is not required,
other systems, which cannot be mapped to free fermions, may equally
well be investigated by means of the DMRG.  The relevance of our
results stands then not only in the accurate test of predicted scaling
relations, and in the determination of nonuniversal coefficients and
corrections beyond scaling, but mostly in the wide applicability of
the method described above to the problem of quantum phase transitions
in disordered systems.  This is a relevant issue in solid-state
physics. For instance, as we discussed in the Introduction, the QIMRF
was obtained as an effective model for the Kondo lattice model.
\cite{gulacsi}  Moreover, recent theoretical developments point out
the relevance of quantum phase transitions \cite{qcp} to the physics
of high-$T_c$ superconductors.  It has indeed been recently suggested
that the properties of these systems close to optimal doping could be
determined by the physics of a nearby quantum critical point,
associated to a magnetic \cite{mag} or an incommensurate
charge-density-wave instability.\cite{htc} The anomalies observed in
the optimally doped metallic phase and the behavior of the critical
temperature have a natural interpretation once the lack of any
characteristic energy scale beside the temperature itself is assumed
as a consequence of the presence of a quantum critical point.

The approach described in this paper allows for a straightforward application
to all those quantum systems that can be mapped onto effective spin models
in the presence of random fields. Our results show indeed that the DMRG
algorithm is a good and reliable alternative to other numerical methods 
within two respects: the possibility of studying systems of a considerable
size, when needed, and the possibility of introducing an external field
coupled to the order parameter, to extrapolate its behavior in the state
of spontaneously broken symmetry.

\acknowledgments
This work was supported by The Swedish Natural Science Research
Council.  S.C.~acknowledges financial support from the Commission of
the European Communities under HCM Contract No. ERBCHBGCT940724.

\end{document}